\documentclass[review]{elsarticle}

\usepackage{lineno,hyperref}
\modulolinenumbers[5]

\journal{Elsearticle}

\begin{document}

\begin{frontmatter}

\title{Tumor stabilization induced by T-cell recruitment fluctuations}

\author{Irina Bashkirtseva}
\ead{irina.bashkirtseva@urfu.ru}
\address{Institute of Mathematics and Computer Sciences, Ural Federal University, 620000, Lenina, 51, Ekaterinburg, Russia}
\author{Lev Ryashko}
\address{Institute of Mathematics and Computer Sciences, Ural Federal University, 620000, Lenina, 51, Ekaterinburg, Russia}

\author{\'Alvaro G. L\'opez}
\address{Nonlinear Dynamics, Chaos and Complex Systems Group, Departamento de  F\'isica, Universidad Rey Juan Carlos\\ Tulip\'an s/n, 28933 M\'ostoles, Madrid, Spain}

\author{Jes\'us M. Seoane}
\address{Nonlinear Dynamics, Chaos and Complex Systems Group, Departamento de  F\'isica, Universidad Rey Juan Carlos\\ Tulip\'an s/n, 28933 M\'ostoles, Madrid, Spain}

\author{Miguel A.F. Sanju\'{a}n}
\address{Nonlinear Dynamics, Chaos and Complex Systems Group, Departamento de  F\'isica, Universidad Rey Juan Carlos\\ Tulip\'an s/n, 28933 M\'ostoles, Madrid, Spain}
\address{Department of Applied Informatics, Kaunas University of Technology, Studentu 50-415, Kaunas LT-51368, Lithuania}
\date{\today}

\begin{abstract}
The influence of random fluctuations on the recruitment of effector cells towards a tumor is studied by means of a stochastic mathematical model. Aggressively growing tumors are confronted against varying intensities of the cell-mediated immune response for which chaotic and periodic oscillations coexist together with stable tumor dynamics. A thorough parametric analysis of the noise-induced transition from this oscillatory regime to complete tumor dominance is carried out. A hysteresis phenomenon is uncovered, which stabilizes the tumor at its carrying capacity and drives the healthy and the immune cell populations to their extinction. Furthermore, it is shown that near a crisis bifurcation such transitions occur under weak noise intensities. Finally, the corresponding noise-induced chaos-order transformation is analyzed and discussed in detail.
\end{abstract}


\end{frontmatter}

\section{\label{sec:level1}Introduction}

It has been recently shown that tumor-immune interactions can
display transient and permanent chaotic dynamics
\cite{Itik10,Msa14}. As discussed in those works, such dynamical
scenarios are relevant when therapies are considered, due to the
resurgence of residual tumor cells that might survive the treatment.
The problem of tumor recurrence is specially important in the
context of immunotherapy \cite{Las19}, since it is related to the
phenomenon of tumor dormancy \cite{Dyn17}. Generally speaking, the
immune system can be regarded as a control system that co-evolves
with a tumor mass acting as a natural selective force. Besides, it
edits its phenotype by selecting those cells that are unresponsive
to immune detection \cite{Agui07,Ten08,Dun04}. As demonstrated in
previous studies \cite{Dyn17}, the levels of T-cell recruitment play
a key role in the maintenance of tumor cell populations at low
levels. The recruitment of effector cells is a very complicated
process, since it involves many different types of cells and
molecules.  The extravasation of leukocytes requires an initial
contact between such cells and the adjacent endothelial cells, which
depends on different adhesion molecules. Once the cells adhere to
the walls of the vessels, the immune cells traverse them through
diapedesis, which again relies on several cytokines. Finally,
chemokines bias their random walks to the tumor site
\cite{Jan12,Tra03}.

Mathematical models of tumor growth and its interaction with the immune system have demonstrated their potential to explain different properties of tumor-immune interactions \cite{Bel00,Wod14,Bel08}. More specifically, a two-dimensional ordinary differential equation model describing the co-evolution of tumors in the presence of an immune response was originally presented by Kuznetsov \emph{et al.} \cite{Kuz94}. This seminal model and their evolved variants have been extensively studied from the point of view of phase space analysis and the theory of bifurcations \cite{Dep03,Val05,Dep06,Val14,Fre00}.

Due to a strong nonlinearity, models of tumor-immune interaction
frequently exhibit multistability with the coexistence of regular
and chaotic regimes. It is well known that a presence of inevitable
random disturbances in nonlinear systems can result in crucial
changes in the dynamical behavior and cause various phenomena, e. g.,
the stochastic resonance \cite{McD08}, noise-induced transitions
\cite{Hor84,Ani07,BNR_CNSNS18}, stochastic excitement
\cite{Lin04,RS_PRE17} and noise-induced chaos \cite{Gao99}. Stochastic
effects in tumor-immune systems have received much attention in the past few years
(see, e. g. \cite{Fia06,Car10,Liu18,Li19,Yang19} and bibliography
therein). But even in the deterministic case, not much research has
been oriented towards the importance of the levels of the T-cell
recruitment \cite{Dyn17} and, as far as the authors are concerned,
none has evaluated in detail the effects of modifying the parameter
that simulates this phenomenon. Furthermore, the specific nature of
this function varies among different studies \cite{Kuz94,Val05} with
little justification. Contrary to the process of tumor cell lysis,
which has been more deeply studied from the mathematical point of
view, the function describing the rate of T-cell recruitment to the
tumor site clearly deserves more attention. As previously commented,
this process is of tremendous complexity and great variations among
different types of tumors are expected, depending on the tissue
location and other factors, as for instance the degree of
inflammation. Moreover, even within a particular tumor that evolves
under the dynamical influence of the adaptive immune response, high
fluctuations of this parameter must occur as time goes by.

In the present work, we study the influence of parametric
perturbations of the T-cell recruitment in dynamical system that contains three ordinary differential equations. After introducing the
unperturbed dynamical system in Sec.~2 and investigating the
bifurcation phenomena therein, we write down the Langevin equations
using multiplicative white Gaussian noise in Sec.~3. We then proceed
to present our main results and characterize the different types of
dynamical transitions observed in the probabilistic model using
standard tools from the theory of stochastic processes. Then, in
Sec.~4, we delve deeper into the noise-induced transition from the
chaotic regime to the globally stable regime. As usual, the closing
section is devoted to expose the most resounding findings and
possible future research concerning the mathematical modeling of
tumor and immune cell recruitment in the present context.

\section{\label{sec:level2} Deterministic mathematical model}

We begin by considering a dynamical model of tumor-immune cell interactions, whose dynamical equations read
\begin{equation}
    \begin{array}{lr}
       \dot{T}=r_{1}T\left(1-\displaystyle\frac{T}{k_{1}}\right)-a_{1 2}T H-a_{1 3}T E \bigskip\\
       \dot{H}=r_{2}H\left(1-\displaystyle\frac{H}{k_{2}}\right)-a_{2 1}H T \bigskip\\
       \dot{E}=r_{3}\displaystyle\frac{E T}{T+k_{3}}-a_{3 1}E T-d_{3}E,
     \end{array}
\end{equation}
where the variable $T(t)$ represents the tumor cells, $H(t)$ describes the healthy host cells that form the tissue where the tumor has emerged and $E(t)$ stands for the effector immune cells ($\emph{e. g.}$ cytotoxic CD8+T lymphocytes). This competition Lotka-Volterra based model is essentially the same as another model originally presented in previous works \cite{Dep03} to study optimal schedules in cancer chemotherapy, and we refer the reader to such references for a more detailed description. All the biological assumptions considered to develop the model equations are based on both accepted knowledge of basic laws governing tumor growth and the immune system function \cite{Kuz94}. Moreover, we recall that some versions of the present model have been tested against \emph{in vivo} experimental data \cite{Dep06, Val14}. The cancer and the healthy cells exhibit limited growth represented by two respective logistic functions, with similar rates of growth and carrying capacities \cite{Val14}. It is also considered that tumor and healthy cells compete for space and resources. Concerning the immune cells, it is assumed that cytokine emission triggers the adaptive immune response producing the recruitment of cytotoxic T-cells to the tumor domain. This phenomenon is considered to act without time-delay and it is approximated by a Michaelis-Menten functional response. Again, a Lotka-Volterra competition term is assumed between tumor and immune cells, since the former find ways to counterattack the latter \cite{Val14}. Finally, after several encounters (or in the absence of competitors), the effector cells inactivate (or die off) at a constant per capita rate $d_{3}$.

A non-dimensionalized model can be readily derived, yielding the simplified equations
\begin{equation}
\begin{array}{l}
\dot x=x(1-x)-a_{12}xy-a_{13}xz\\
\dot y=r_2y(1-y)-a_{21}x y\\
\dot z=r_3 \displaystyle\frac{x}{x+k_3}z-a_{31}xz-d_3z,\\
\end{array}
\label{1}
\end{equation}
where, $x(t)$, $y(t)$ and $z(t)$ are densities of populations of
tumor cells, healthy host cells, and effector immune cells,
respectively. In other works only equilibrium regimes were
covered \cite{Dep03}. On the other hand,
the constant input of immune cells into the tissue can be eliminated,
disregarding the innate immune response \cite{Itik10}. Parametric zones
with regular and chaotic oscillatory dynamics were discovered and
extensively described from a mathematical point of view. Later on, a
new parametric zone, closer to the original set of parameters was
found, with coexisting chaotic and equilibrium attractors near a
boundary crisis \cite{Msa14}. Here we follow this last reference and
fix the parameters values as
$$ a_{12}=0.5,\; a_{21}=4.8,\; a_{13}=1.2,\; a_{31}=1.1,\;r_2=1.2,\; d_3=0.1,\; k_3=0.3$$
in order to study the system dynamics under the variation of the
parameter $r_3$. This parameter is relevant in our work since it characterizes the rate of immune cell recruitment as a consequence of their interaction with the tumor cells and
subsequent cytokine emission. As a reminder, this model parameters
are are reasonable from an empirical point of view \cite{Val14}. Two different parameters in comparison with the original set of values
correspond to $r_{3}$ and $a_{21}$, which take higher values. This set of values corresponds to very immunogenic and aggressively competing tumors (\emph{e. g.} acidification is high).
\begin{figure}
\centering
\includegraphics[width=0.42\textwidth]{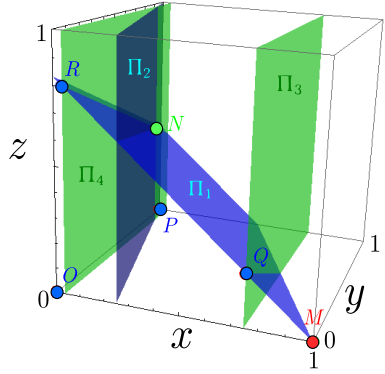}
\caption{\textbf{General view of the phase space}.
The deterministic model represented in Eqs.~(\ref{1}) presents six fixed points, which are the
result of different intersecting planes. In addition to the $x$, $y$
and $z$ planes (not colored), we can see four more nullclines
$\Pi_{i}$, with $i=1,...,4$. The equilibrium representing the
malignant state $M$ is represented in red color, while another fixed
point $N$ taking place for small tumor size populations is plotted
in green. Chaotic and periodic oscillations organize around this
unstable spiral.} \label{fig1}
\end{figure}

\begin{figure}
\centering
\includegraphics[width=1.1\textwidth]{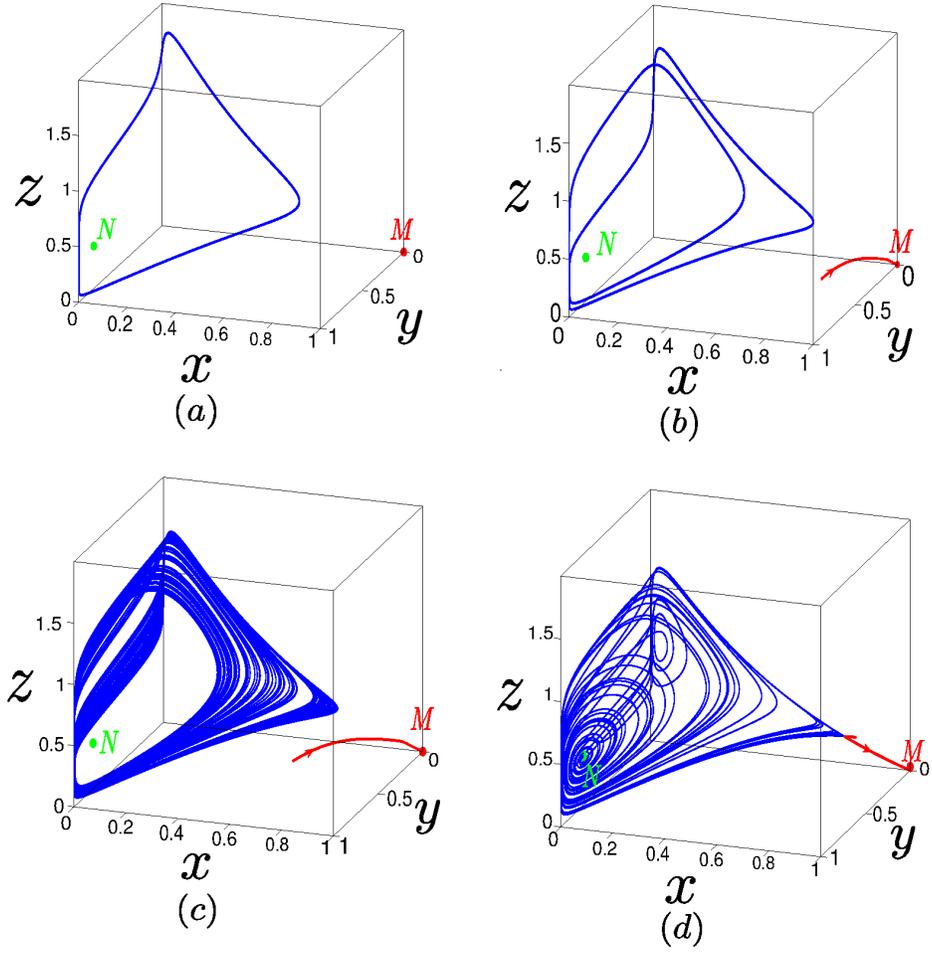}

\caption{\textbf{Regular and chaotic oscillatory attractors}. The phase space trajectories of the model appearing in Eqs.~(\ref{1}) are represented. A saddle fixed point $N$ and a malignant attractor $M$ appear in green and red color, respectively.  (a) A periodic cycle is found for $r_3=1.6$.
(b) A two period trajectory appears for $r_3=1.5$. (c) A chaotic
attractor formed by two bands is found for $r_3=1.44$ (e) A chaotic
attractor results for the parameter value $r_3=1.291$, which is close
to a boundary crisis.} \label{fig2}
\end{figure}

As we have previously indicated, the parameter $r_3$
plays a key role in the present study \cite{Bif17}. Small changes
of $r_3$ can result in catastrophic shifts, even for the
deterministic dynamical system appearing in Eqs.~(\ref{1}). These shifts
can be explained mathematically by corresponding bifurcations. In
this study, we will consider $r_3\in[1.25,1.65]$. In this range of
parameters, the dynamical system presents five or
six fixed points in the positive octant as shown in the 3D phase
space depicted in Fig.~\ref{fig1}. For our numerical
simulations, we take the value $r_3=1.291$, again according to
Ref.~\cite{Msa14}. Using this parameter value the origin $O(0,0,0)$ is a saddle fixed point. It has two positive eigenvalues aligned with the $x$ and $y$ axes, while a third negative eigenvalue is associated to the remaining $z$-axis. The point
$P(0,1,0)$ represents the tumor-free solution, for which only
normal cells can be found. As long as we keep in the plane $x=0$, this fixed point is stable. On the contrary, the point $M(1,0,0)$ represents the stable solution for which only tumor cells can be found. Therefore, this third fixed point represents a dangerous malignant attractor. Indeed, for this equilibrium the eigenvalues of the Jacobian matrix are
$$\lambda_1=-1,\;\;\;\lambda_2=r_{2}-a_{21}=-3.6,\;\;\;\lambda_3=\frac{r_3}{1+k_3}-a_{31}-d_3.$$
This equilibrium is always stable in the plane $z=0$. In the 3D phase
space, this equilibrium is stable for
$r_3<r_3^*=(a_{31}+d_3)(1+k_3)=1.56$. Thus, if the recruitment of the T-cells is high enough, the malignant state loses its stability. The saddle fixed point $Q(0.75,0,0.21)$ has a two-dimensional stable manifold that separates the basins of attraction of the malignant tumor fixed point and the chaotically oscillating attractor. Two spiral-saddles $R(0.04,0,0.8)$ and $N(0.04,0.85,0.45)$ remain. The former $R$ spirals stably, with its unstable direction pointing closely to the direction of the $y$-axis. The later displays inverted stability, \emph{i.e.}, it is the spiral that is unstable. These two confronted spiral-saddles ceaselessly interchange their dynamics leading to the heteroclinic chaotic (or periodic) behavior of this dynamical system (see Fig.~\ref{fig2}). The non-degenerate equilibrium $N(\bar x,\bar
y,\bar z)$ has positive coordinates
$$\bar x=\frac{r_3-a_{31}k_3-d_3-\sqrt{(r_3-a_{31}k_3-d_3)^2-4 a_{31} d_3 k_3}}{2 a_{31}},$$
$$
\bar y=1-\frac{a_{21}\bar x}{r_2},\quad\bar z=\frac{1-\bar x-a_{12}\bar y}{a_{13}}.$$

As depicted in Fig.~\ref{fig2}(a), for $r_3=1.65$, the additional (to $M$) attractor $\cal{A}$ is regular and periodic. This attractor is a stable limit cycle. At the bifurcation point $r_3=1.573$, this 1-cycle loses its stability and is transformed into a stable 2-cycle. A further decrease of the recruitment produces a standard period-doubling cascade, and the attractor $\cal{A}$ is transformed into the full chaotic strange attractor shown in Fig.~\ref{fig2}(c).

\begin{figure}
\centering
\includegraphics[width=0.95\textwidth]{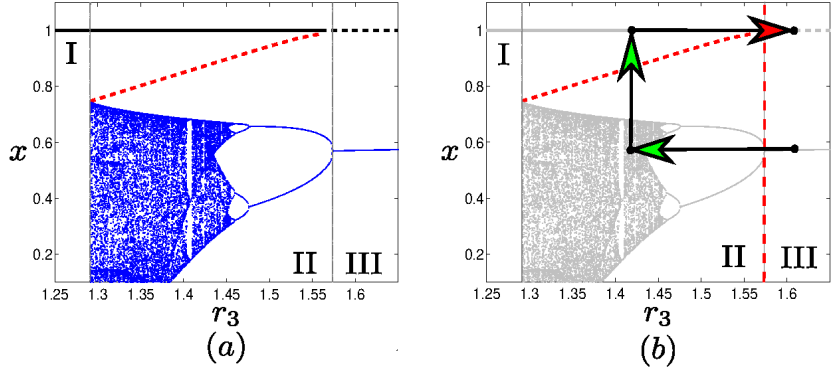}
\caption{\textbf{Bifurcation diagram of the deterministic system}. (a) The tumor size population is plotted against the parameter $r_{3}$. The equilibrium $M(1,0,0)$ is plotted with a black line (solid for stable and dashed for unstable). The oscillatory attractor $\cal{A}$ is plotted in blue. It loses stability and leads to chaotic behavior as the recruitment decreases through a period doubling cascade. The separatrix is shown in red. (b) A hysteresis loop showing how the fluctuations move along the bifurcation diagrams. Green arrows are allowed transitions, while the red arrow represents a forbidden transition.}
\label{fig3}
\end{figure}

Thus, the attractor ${\cal A}$ represents a variety of stable dynamical coexistence of healthy, immune and tumor cells populations, with changing $r_3$. The chaotic attractor disappears at the crisis bifurcation point $r_3^c=1.2909$. The system behavior near this crisis bifurcation point was studied in detail in \cite{Msa14}. In the bifurcation diagram appearing in Fig.~\ref{fig3}, the equilibrium $M(1,0,0)$ is shown by a black line (solid for stable and dashed for unstable), while a family of attractors $\cal{A}$ shaping a Feigenbaum's tree is plotted in blue color. Here, $x$-coordinates of local maxima of attractors are shown. In summary, in the interval $1.25\leq r_3 \leq 1.65$ one can distinguish three parametric zones, depending on the levels of T-cell recruitment:

\begin{enumerate}

\item A \emph{dead} monostable zone ($1.25\leq r_3 \leq r_3^c$): here, the equilibrium $M(1,0,0)$ is a single global attractor and the solutions of the deterministic system tend towards $M(1,0,0)$ asymptotically for any initial data. Therefore, in this range, the population of effector and healthy cells are out-competed by tumor cells independently of the relative initial sizes of the cell populations. In other words, insufficient levels of recruitment permit the settlement of the tumor no matter what.

\item A \emph{hazardous} bistable zone ($r_3^c\leq r_3 \leq r_3^*$): in this region, depending on the initial conditions, solutions of Eqs.~(\ref{1}) may tend either to $M(1,0,0)$ or to the oscillatory attractor ${\cal A}$ (chaotic or periodic). In Fig.~\ref{fig3}, the separatrix dividing basins of attraction of $M$ and ${\cal A}$ is shown by a red dashed line. There is here enough recruitment of effector cells so as to hinder the tumor growth, as long as the size of the latter is not too close to its carrying capacity.

\item A \emph{safe} monostable zone ($r_3^*\leq r_3 \leq 1.65$): here, all solutions starting from the initial point with positive coordinates tend to the periodic attractor. In the narrow subinterval $1.56<r_3<1.573$, this attractor is a 2-cycle, and in the subinterval $1.573<r_3<1.65$ this attractor is a 1-cycle. The recruitment here is more than sufficient to keep the tumor below its carrying capacity, performing periodic oscillations in size.

\end{enumerate}

Examples of oscillatory attractors are shown in Fig.~\ref{fig2} in blue color. In red color, phase trajectories leaning towards the equilibrium $M(1,0,0)$ are plotted. As can be seen in the bifurcation diagram appearing in Fig.~\ref{fig3}, this system exhibits a plethora of bifurcations and dynamical regimes, both regular and chaotic. Furthermore, depending on the parameter $r_3$, monostable and bistable zones can be determined. It is well known that noise represents an inevitable attribute in the dynamics of cell biology that can crucially change the fate of the system. In the following lines, we examine how noise impacts the interactions of tumor, healthy and immune cells populations in mono and bistable zones. As we shall see, recruitment fluctuations can dramatically impair the establishment of the tumor.

\begin{figure}
\centering
\includegraphics[width=1.0\textwidth]{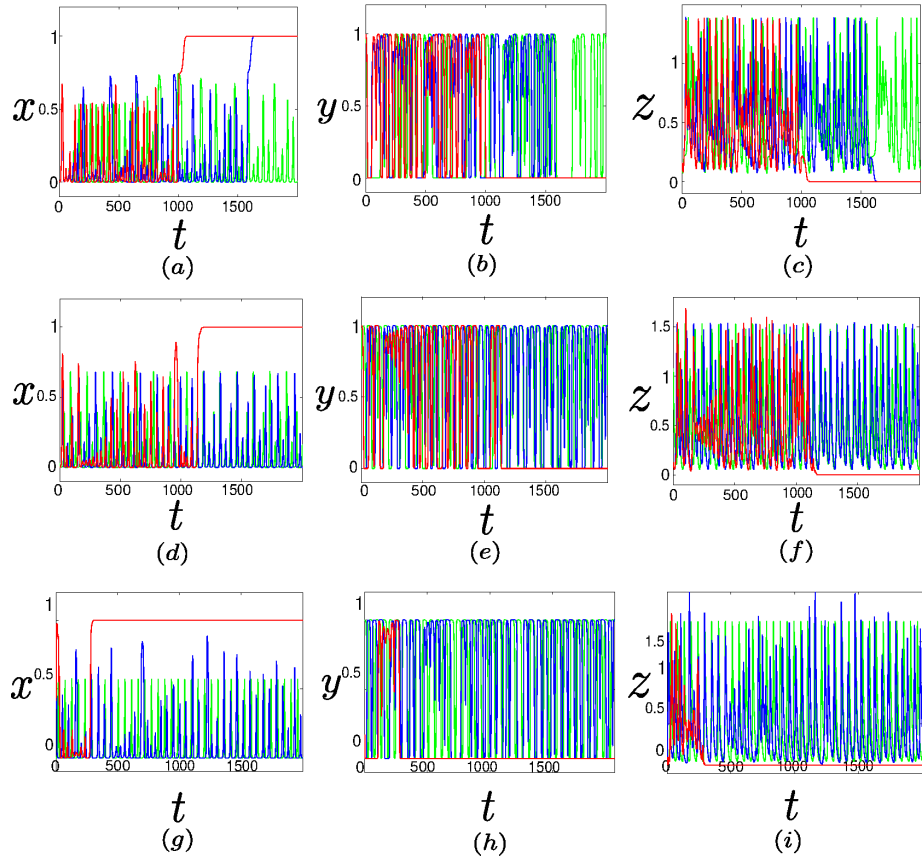}
\caption{\textbf{Time series of the solutions of both the deterministic (green color) and the stochastic system (red and blue colors)}. (a-c) The evolution of tumor $x$, healthy $y$ and immune cells $z$ for a recruitment value of $r_3=1.291$ in the \emph{hazardous} zone and noise intensities $\varepsilon=0.01$ (blue) and $\varepsilon=0.02$ (red). (d-f) A higher recruitment value in the same parameter region with $r_3=1.409$ and noise intensities $\varepsilon=0.02$ (blue) and $\varepsilon=0.2$ (red). (g-i) A recruitment value of $r_3=1.6$ and noise intensities $\varepsilon=0.4$ (blue) and $\varepsilon=0.8$ (red).}
\label{fig4}
\end{figure}

\section{Stochastic model}

Now we consider the behavior of the system of Eqs.~(\ref{1}) under random disturbances since the tumor growth does not usually take place in a deterministic manner. To simulate the effect of external fluctuations, we use the following system of stochastic differential equations in the Ito's sense
\begin{equation}\label{2}
\begin{array}{l}
\dot x=x(1-x)-a_{12}y x-a_{13}z x\\
\dot y=r_2y(1-y)-a_{21}xy\\
\dot z=\left(r_3+\varepsilon \xi(t)\right)\displaystyle\frac{x}{x+k_3}z-a_{31}x-d_3.
\end{array}
\end{equation}
Here, the random forcing  $\xi(t)$ is an uncorrelated white Gaussian noise term with parameters $\langle\xi(t)\rangle=0,\;\langle\xi(t)\xi(\tau)\rangle=\delta(t-\tau),$ and $\varepsilon$ represents the noise intensity.  Since the model is dissipative, we can neglect small correlations in the noise. Moreover, because we are focusing our attention on the effect of fluctuations in T-cell recruitment, the noise term has been introduced in a multiplicative way by a perturbation of the recruitment parameter $r_3$. We also recall that additive noise is inadequate in population models, since cell populations are restricted to take positive values. Consequently, we assume that the fluctuations in the immune cell populations are dominantly produced through the complex dynamical process of cell recruitment described in the introduction, which depend on multiple internal events, but also on events that are external to the tumor. This fact justifies the use of a Langevin approach \cite{Van92}. Note that the equilibrium $M(1,0,0)$ of the model appearing in Eqs.~(\ref{1}) is also the equilibrium of the stochastic model represented by Eqs.~(\ref{2}), because of the structure of the noisy term. Therefore, the solutions of the former starting from $M(1,0,0)$ never leave this point for any $\varepsilon$. This fact will be crucial to explain the dynamics of the system and will be responsible of producing a previously undetected extinction effect.

We begin by analyzing the stochastic dynamics of the model appearing in Eqs.~(\ref{2}) in the \emph{hazardous} bistable zone $r_3^c\leq r_3 \leq r_3^*$, since it is the most sensitive to noise of the two last parameter regions. Under time varying random disturbances, solutions can leave the deterministic 
oscillatory attractor ${\cal A}$. For weak noises, random trajectories are concentrated near the
deterministic attractor. As noise increases, the dispersion of
random trajectories increases as well. Now the random trajectory can
cross the separatrix between basins of attraction of $M$ and ${\cal
A}$, to later on approach the point $M$. The density $x$ of tumor
cells tends to one, and the healthy cell populations of the tissue
$y,$ $z$ are lead to extinction. Here, we observe a noise-induced transition from the regime of the coexistence of healthy, immune and tumor cells to the absolute tumor dominance.
Importantly, this extinction of both cell compartments occurs in an irreversible fashion, since once the cytotoxic effector cells are eradicated, no more T-cells can be recruited. We insist that this
occurs in our model because the recruitment process is based on a feedback mechanism relying on the existence of effector T-cells. The same holds for the healthy normal cells, which can only grow from
other cells (the stem ones) of the same lineage.

\begin{figure}
\centering
\includegraphics[width=0.9\textwidth]{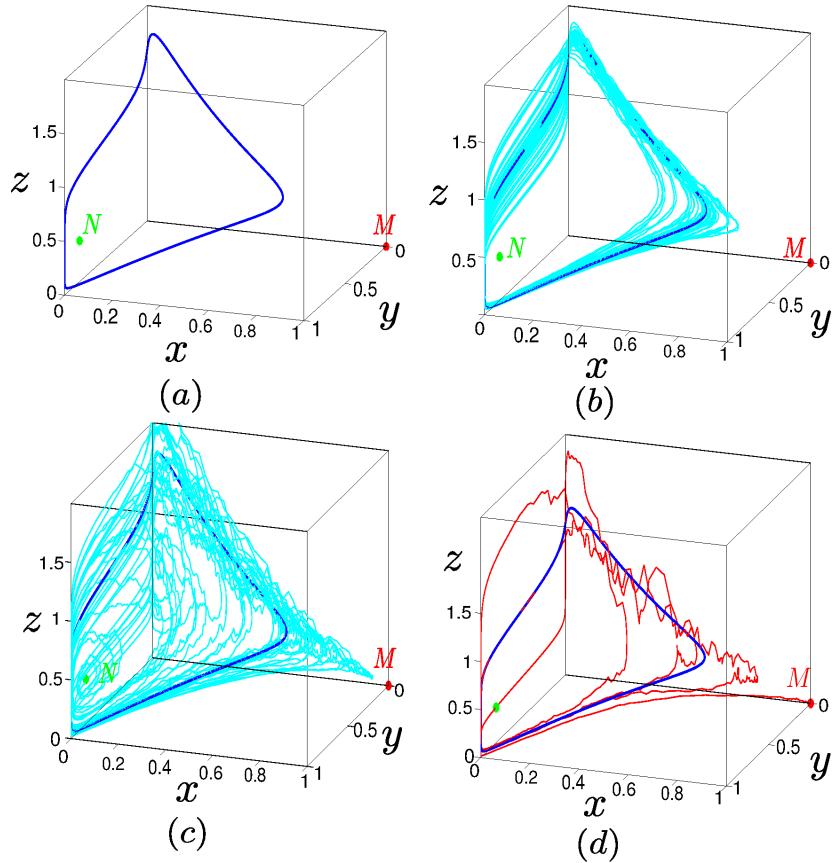}
\caption{\textbf{The effects of noise on phase space trajectories}. (a) The deterministic periodic cycle for $r_{3}=1.6$. (b) The same trajectory with a white multiplicative noise $\varepsilon=0.1$, which is not sufficient to break the periodicity of the cycle. (c) The same situation with higher fluctuations given by $\varepsilon=0.2$. Now the trajectories resemble the deterministic chaotic attractor previously shown, with fluctuations superimposed. However, the malignant attractor is not easily reached. (d) When the fluctuations are sufficiently intense ($\varepsilon=0.4$), the tumor is rapidly stabilized.}
\label{fig5}
\end{figure}

The effect of a noise-induced extinction is frequently observed in biological systems \cite{Sar18,Tsi14}, and it is not surprising at all in multistable systems.
Such stochastic transitions are numerically shown in Fig.~\ref{fig4}
by inspection of the time series for different values of $r_3$.
As is immediately observed, the closer $r_3$ it is the crisis
bifurcation point $r_3^c$, the smaller the size of the perturbations
required to transfer the system from the oscillatory attractor $\cal
A$ to the equilibrium $M$. The dynamics of trajectories in phase
space are shown in Fig.~\ref{fig5}. More importantly, under the
influence of noise, such noise-induced transitions can be observed
from the monostable zone $r_3^*\leq r_3 \leq 1.65$, which was
classified as \emph{safe} in the deterministic system. Indeed, in this
zone (see Fig.~\ref{fig2}(d) for $r_3=1.6$), increasing the
fluctuations results in the transition of the stochastic system to
the equilibrium $M$. Note that in this zone the equilibrium $M$ is
unstable. Therefore, this phenomenon can be described mathematically
as a noise-induced stabilization of the unstable equilibrium by
parametric noise.

The process occurs through a hysteresis phenomenon which, as far as
we know, it had not been previously reported in the literature on mathematical oncology and physics of cancer. It can be explained in the bifurcation diagram as 
shown in Fig.~\ref{fig3}(b). In the first place, recruitment fluctuations permit an 
excursion from the \emph{safe} region to the \emph{hazardous} zone. This phenomenon produces a instability of the
tumor cell population trajectories, which are now allowed to perform
chaotic oscillations, occupying a bigger portion of the phase space.
But now, since the fluctuations in the cell recruitment are still
present, a transition similar to the one described in the previous
paragraph might occur. Then, since the fluctuations are not
affecting anymore the extincted populations, they can not drive back
the tumor to small values. Collaterally, the healthy cells are also
dragged to extinction and the tumor is stabilized at its maximum
value, making useless the efforts of the cell mediated immune
response. This fact suggests that the process of recruitment in
aggressively growing tumors might be very important concerning the
effectiveness of the immune response and the associated preservation of
the healthy host tissue.

\section{Noise-induced and chaos-order transitions}

For the detailed study of such noise-induced transitions, we will use a parametric analysis of some statistics. First, we consider the mean values $\mu_x, \mu_y,$ and $\mu_z$ of the coordinates of random solutions of system (\ref{2}) during the time interval $ [0,T]$. For example, the time average of the tumor cell population can be written as $$ \mu_x=\frac{1}{L}\sum_{k=0}^L x(kh),\qquad h=\frac{T}{L}.$$ Another important statistic is the size of the fluctuations, which can be estimated by means of the second cumulant, defined as $$ \sigma_x^{2}=\frac{1}{L}\sum_{k=0}^L x^{2}(kh)-\mu^{2}_{x},\qquad h=\frac{T}{L}.$$

\begin{figure}
\centering
\includegraphics[width=1.0\textwidth]{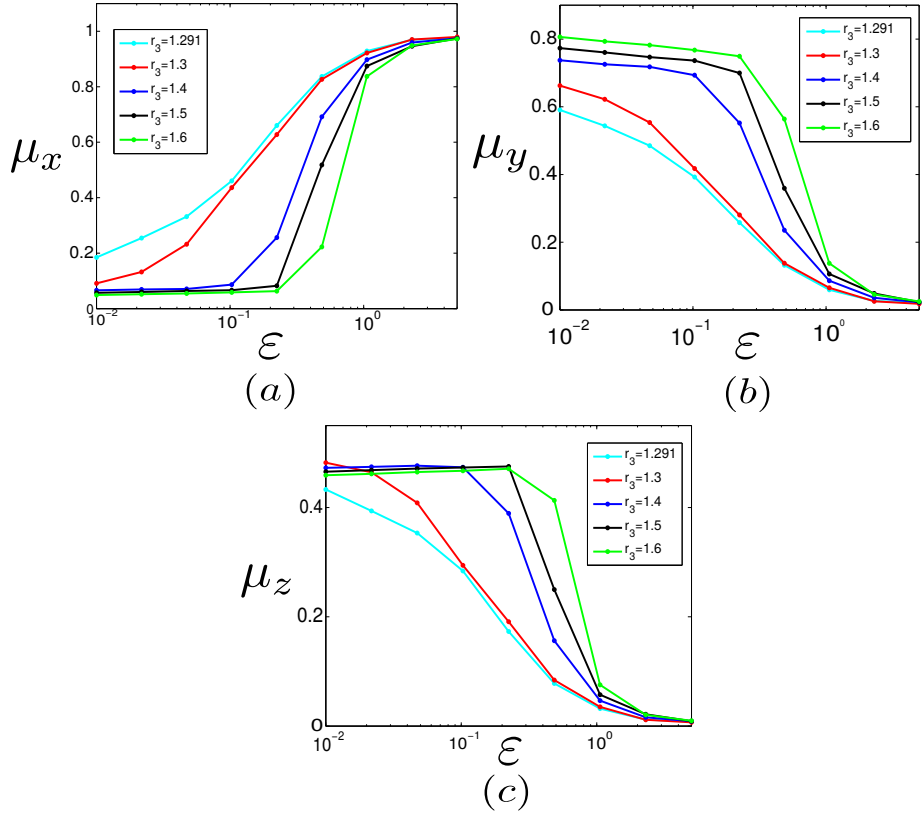}
\caption{\textbf{The time-averaged values of the solutions of the stochastic system for the three cell populations}. Here we are using an interval $[0,T]$ with $T=1000$.(a) Average size of tumor cells. We see a transition from small tumor burdens to sizes close to the carrying capacity. This means that noise facilitates tumor stabilization.  (b) Average size of healthy cells. (c) Average size of immune cells. The $\varepsilon$-interval where the transition takes place shrinks as $r_{3}$ increases. The pattern for normal and immune cells is complementary to tumor cells.}
\label{fig6}
\end{figure}

As an initial value of these solutions, we take the unstable non-trivial equilibrium $N$. A dependence of the mean values on the parameters $r_3$ and $\varepsilon$ is shown in Fig.~\ref{fig6} for $h=0.01$ and $\;T=10^3$. As can be seen, under increasing noise intensity, the mean value $\mu_x$ of tumor cells monotonously increases and tends to $\mu_x = 1$, whereas the mean values $\mu_y$ of healthy cells and $\mu_z$ of immune cells tend to zero. This scenario is observed in the bistable zone (see $r_3=1.291, 1.3, 1.4$) and in the monostable zone ($r_3=1.6$) as well. Note that mean values sharply change in a rather narrow window ($\varepsilon$-interval), specially for high values of the recruitment. Therefore, we see that those situations in which the immune response is weaker, the transition due to fluctuations occurs more easily and progressively for increasing values of the fluctuations.

\begin{figure}
\centering
\includegraphics[width=1.0\textwidth]{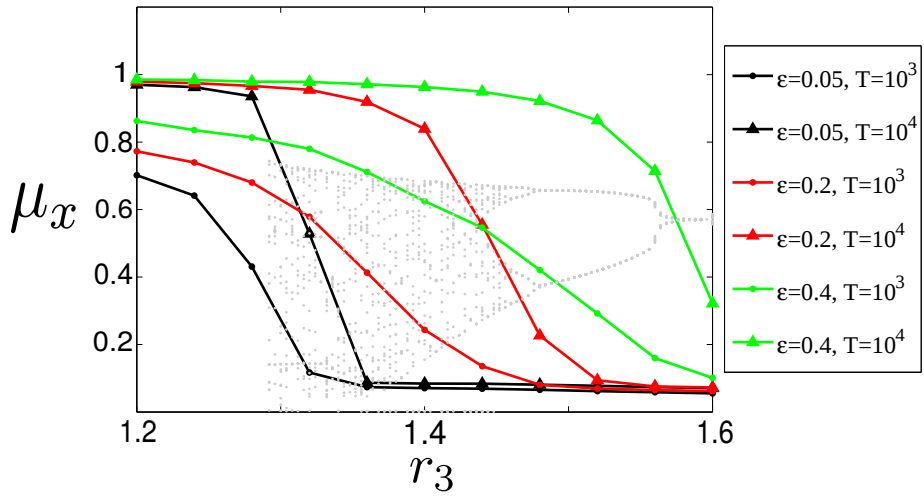}
\caption{\textbf{Mean values $\mu_x$ of the solutions of the stochastic system against the recruitment of T-cells}. Two time intervals $T=10^3$ (circles) and $T=10^4$ (triangles) are considered, and three different noise intensities $\varepsilon=(0.05,0.2,0.4)$, using different colors. A dependence on time average is detected. The longer the time interval, the sharper the transition from tumor stabilization to tumor contention. These suggests that, if we let $T$ approach to infinity, in most situations we obtain a binary dynamical situation under the effects of noise. The bifurcation diagram is shown in the back in light gray to better correlate the dynamics and the mean values.}
\label{fig7}
\end{figure}

From a mathematical point of view, the previous facts occur because the separatrix curve is closer to the chaotic attractor and this facilitates the transition as shown in Fig.~\ref{fig3}. Therefore, the location and the width of this $\varepsilon$-interval depends strongly on $r_3$. The higher the levels of T-cell recruitment are, the greater it is the noise required for a transition to the \emph{dead} zone, and the more sudden is the appearance of this transition. Note that the green curves reflect a variation of mean values for $r_3=1.6$, which belong to the \emph{safe} zone. As one can see, the deterministic instability of the \emph{dead} equilibrium $M$ does not guarantee a survival of healthy
cells in presence of noise if perturbations are present. This fact can be interpreted as a noise-induced
extension of the dangerous zone. In Fig.~\ref{fig7} the previous
conclusion is confirmed by studying the mean values of $\mu_x$ as a
function of $r_3$ for the random solutions starting from the point
$N$ for different noise intensities $\varepsilon$ and observation
times $T$. Here, the extremum values of the $x$-coordinates of the
attractor $\cal A$ of the deterministic system of Eqs.~(\ref{1}) are also
shown in grey. In this figure, one can compare how the mean values
$\mu_x$ depend on $r_3$ for the two values $T=10^3$ and $T=10^4$.

\begin{figure}
\centering
 \includegraphics[width=1.0\textwidth]{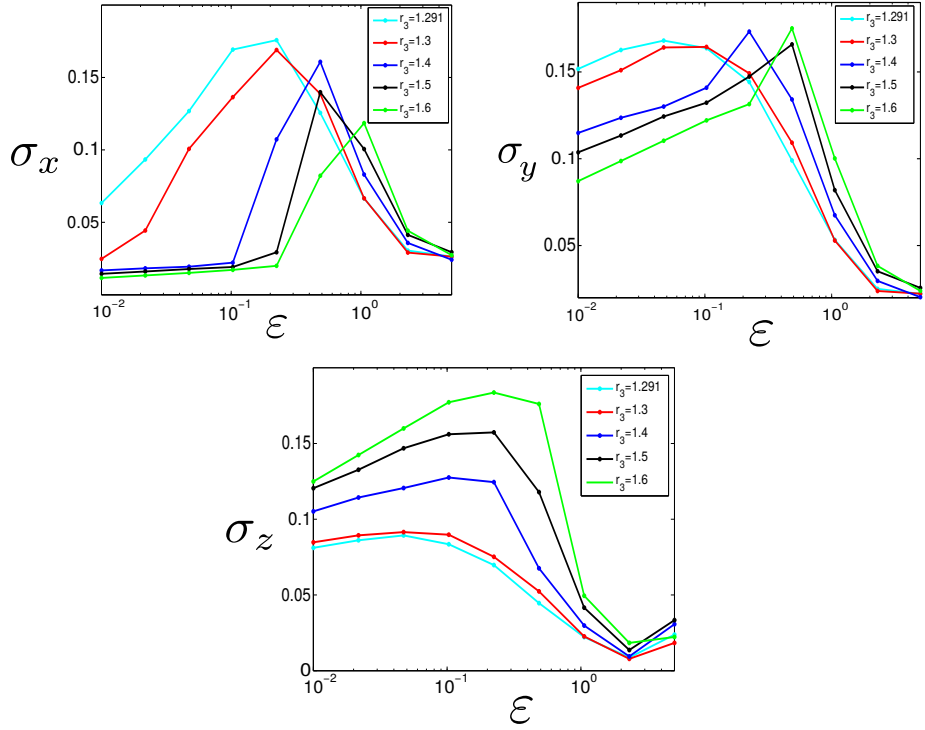}
\caption{\textbf{Time averaged values for the mean squared deviation of the size of the three cell populations.} Again, we are using an interval $[0,T]$ with $T=1000$. The fluctuations peak around values for which the mean values of the cell populations take intermediate values. In other words, for medium noise intensities we see the most fluctuating cell populations, and the more uncertainty we have on the asymptotic state of the system.}
\label{fig8}
\end{figure}

\begin{figure}
\centering
\includegraphics[width=1.0\textwidth]{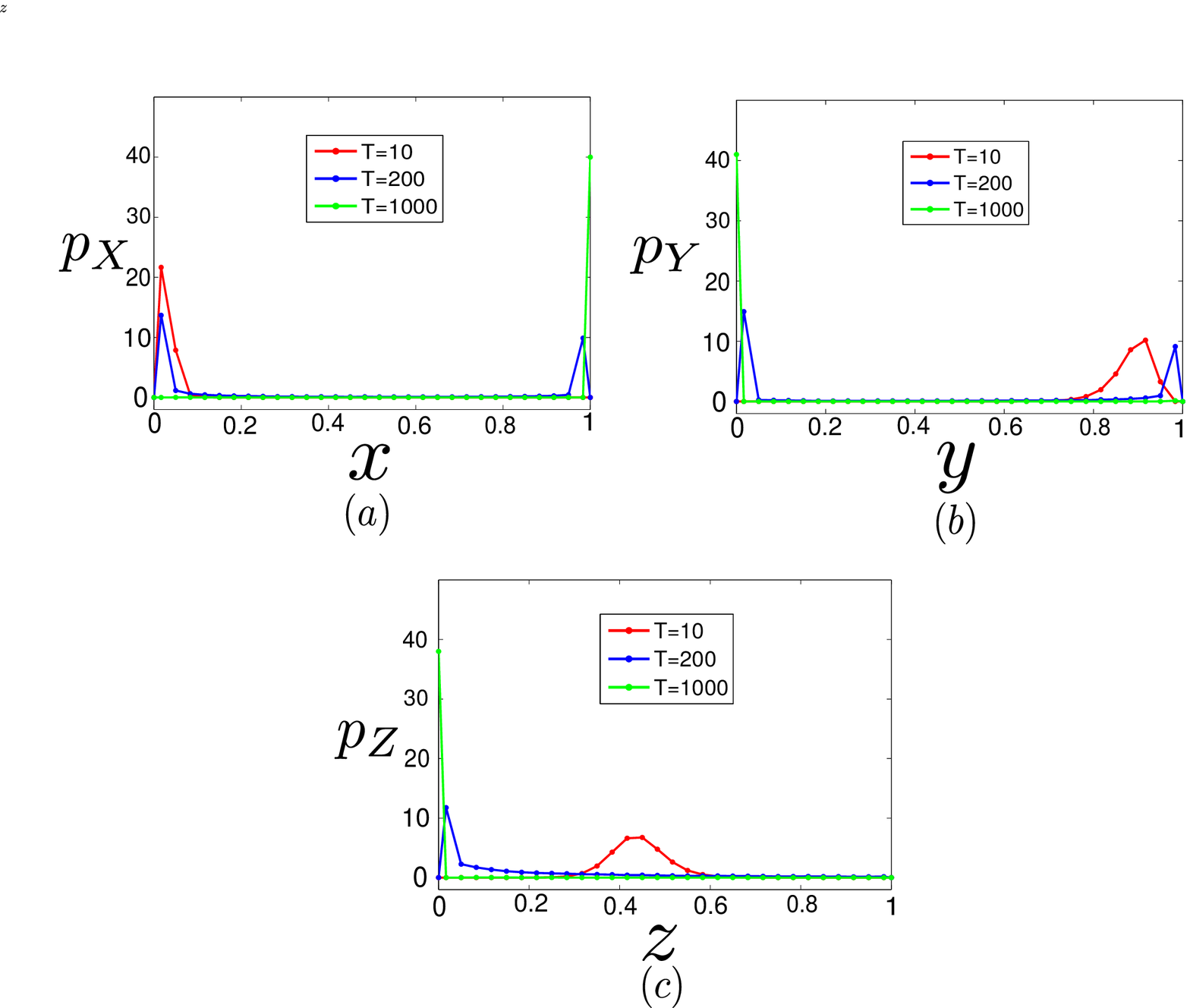}
\caption{\textbf{Evolution of marginal probability density of each cell population with time.} (a) The
probability density $p_{X}(x)$ of tumor cells at three time instants
$T$. The probability distribution, initially placed at the fixed
point $N$, starts to diffuse and later on splits into two. Finally
it concentrates at the malignant attractor $M$. (b) The probability
density of healthy cells $p_{Y}(y)$ at three time instants $T$. The
evolution is opposed to the evolution of the tumor cells, but
dynamically similar. (c) The immune cell's probability density
$p_Z(z)$ rapidly diffuses and drifts towards a delta-peaked
distribution representing extinction.} \label{fig9}
\end{figure}

Along with the mean values, the time averaged size of the
fluctuations of each cell population has been
computed and it is depicted in Fig.~\ref{fig8}. For the tumor cell
populations the size of the fluctuations $\sigma_x$ are peaked at
intermediate values of the sigmoid curves describing. Certainly,
this occurs because the fluctuations are higher between the two
limiting situations. The former corresponds to a tumor cell
population that is retained at a stable small burden by the immune
cells, and the other represents the opposite case, in which tumor
dominates the struggle. Exactly the same occurs for the fluctuations
of the other cell populations, since they are always in a relation
of dependence. For the sake of completeness, we have also shown how
the probability densities evolve as times goes by. This study is
tantamount to deriving the Fokker-Planck equations, but
computationally it is much more affordable. As can be seen in
Fig.~\ref{fig9}, an initial delta-peaked distribution of the tumor
size localized at a point very close to $N$, evolves in time by splitting into two peaks.
Then, this bimodal transient distribution evolves back into a sharp
unimodal distribution centered at $x=1$. Exactly the same process is
observed for the normal cells, but with opposite fate. For the
immune cells, however, the distribution directly evolves towards a
sharp peak at zero.

We now proceed to characterize the probability of the transition of the stochastic model from the stable oscillatory regime of coexistence of tumor, healthy, and immune cells, to the \emph{dead} zone. To describe probabilistically such transition, we need to introduce some formalization. Here, we assume that this system falls into the \emph{dead} zone when $x(t)$ becomes close to one. Let $\delta$ be a parameter representing the proximity to one, then we define the region $M_{\delta}$ for which $x\geq \tilde x=1-\delta$ as a threshold value for $x$-coordinates of random solutions. In our calculations, we take $\delta = 0.01$. This means that the stochastic differential equation appearing in Eqs.~(\ref{2}) falls into the \emph{dead} zone when $x(t)>\tilde x=0.99$. Under this assumption, a conditioned probability $p(M_{\delta}|N)$ of the transition of the stochastic system with the noise intensity $\varepsilon$ to the \emph{dead} zone during the fixed time interval $[0,T]$, when the system is originally at $N$. This magnitude represents the probability of the event $x(t)>0.99$ for $0\leq t\leq T$ with initial conditions as given.

\begin{figure}
\centering
 \includegraphics[width=0.6\textwidth]{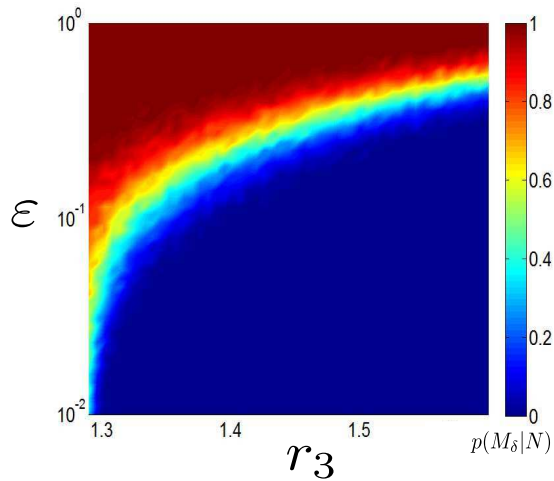}
\caption{\textbf{The probability of exit towards the \emph{dead} zone}. The \emph{dead} zone is defined as $M_{\delta}=\{x : x>1-\delta\}$, with $\delta=0.01$ for the stochastic system conditioned to $N$, as a starting point. The
color bar represents the probability of that event. As can be
appreciated, a narrow $\varepsilon$-interval is observed for the
transition to take place. Importantly, this window of
transition describes a nonlinear curve as $r_{3}$ is varied.}
\label{fig10}
\end{figure}

In Fig.~\ref{fig10}, a colormap of $p(M_{\delta}|N)$ is plotted in the parameter space $(r_3,\varepsilon)$, with a fixed value $\delta=0.01$. As can be seen, this function sharply increases in a very  narrow $\varepsilon$-interval. A location of this $\varepsilon$-interval defines a threshold noise intensity corresponding to the transition to $M$. It should be noted that the $p(M_{\delta}|N)$ increases rather abruptly as $r_3$ increases. However, the location of this transition as a function of $r_{3}$ follows a nonlinear relation.

\begin{figure}
\centering
\includegraphics[width=0.7\textwidth]{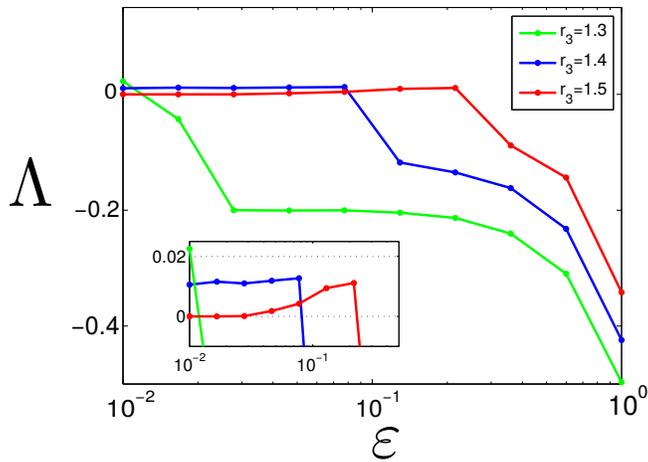}
\caption{\textbf{The largest Lyapunov exponent plotted against the intensity of the fluctuations.} For most values of $r_{3}$ a chaos to order transition is observed as fluctuations increase. However, for a value corresponding to periodic regions close to the period doubling bifurcation, a new phenomenon is observed. In the beginning we have regular dynamics, then the noise produces chaos, and finally the tumor is stabilized. The inset shows a magnification of the region where the Lyapunov exponent grows with the size of the fluctuations.}
\label{fig11}
\end{figure}

To conclude our investigation, we consider how the noise-induced transitions from the regime of the coexistence of healthy, immune, and tumor cells to the tumor dominance are accompanied by a chaos-order transformation. In Fig.~\ref{fig11}, a dependence of the largest Lyapunov exponents $\Lambda$ on the noise intensity $\varepsilon$ is shown for various $r_3$ from the \emph{hazardous} zone. Here, for $r_3=1.3$ and $r_3=1.4$, the deterministic attractor $\cal A$ is chaotic, and for $r_3=1.5$ is regular (2-cycle). A change of the sign of $\Lambda$ from plus to minus marks the transition from chaos to order. As one can see, for $r_3$ close to $r_3^c$ (see green curve for $r_3=1.3$), this transformation is observed for small noise intensities. For $r_3=1.4$ (see blue curve), this transformation from chaos to order occurs for a larger noise intensity.

If we calculate $\Lambda$ for $r_3$ where regular oscillations are observed, then the scenario of these transformations changes. Indeed, for regular limit cycles, the largest Lyapunov exponent for $\varepsilon=0$ equals zero. Under increasing the noise intensity, at first $\Lambda$ increases and becomes positive (see red curve for $r_3=1.5$). With a further increase of $\varepsilon$, the largest Lyapunov exponent begins to decrease and becomes negative. So, here we have detected ``order-chaos-order" transitions, which are induced by noise changes.

\section{Conclusions and Discussion}

We have studied the influence of parametric random fluctuations of T-cell recruitment in a known deterministic 3D cancer model. For parameter values representing aggressively growing tumors that are considerably immunogenic, three subintervals with qualitatively different dynamics were determined. The borders of these zones are defined by the crisis bifurcation of the chaotic attractor $\cal A$ and the loss of the stability of the equilibrium $M$. To study stochastic effects, we considered a forced variant of the system with random fluctuations of the parameter $r_3$. By direct numerical simulation of the time series it was shown that in this system noise-induced transitions to the \emph{dead} zone occur in a wide range of the parameter values of $r_3$.

An important finding of the present work is that fluctuations in the recruitment of T-cells can drive a healthy host tissue to extinction and produce complete tumor dominance. We believe that this effect is interesting from a more general ecological point of view, since we are using a competition Lotka-Volterra based model. The key point is as follows. Consider that two competing species are in a situation of imbalance, in such a manner that, in the absence of any other species, the dominant tends to produce the extinction of the weaker. Now, assume that an equilibrium among the two is kept by a third assistant species that provides balance by destruction of the dominant individuals, leading to chaotic dynamics and bistability. Then, fluctuations which in principle only affect the size of the assistant species can be amplified by chaotic dynamics and lead to the extinction of the weakest individuals. Even if chaos is not at play, as we have shown, this extinction can be induced by the fluctuations themselves, as long as the fluctuations are sufficiently intense. In this manner, the dependence of the weak species on the assistant species becomes very strong, to the point that their fitness is mostly surrogated to their collaborators.

Concerning cancer therapy, the present research sheds light into the importance of the phenomenon of T-cell recruitment. Even though a particular immunotherapy (e.g. CTLA-4 activity modulation by means of monoclonal antibodies \cite{Rib05}, or the blocking of the PD-1 receptor) might lead to a substantial destruction of a tumor by the activation of T-cells, high levels of recruitment over a sustained period of time might be required to obtain the desired tumor remission. Otherwise, fluctuations in T-cell numbers could lead to undesired and unexpected effects, as shown in the present work. To conclude, it is important to note that the effect of fluctuations on tumor and healthy cell populations has been disregarded in the present model. Noise is permanently present in the growth of a tumor through pH variations, temperature, growth factors fluctuations, vascular evolution, etc \cite{Han11,Wei13}. Therefore, further research is deserved to acquire a more complete view of the effect of random perturbations in tumor-immune interactions. In particular, we remark those concerning the intervention of other immune cell lineages and the role of the humoral immune response as well.

\section{Acknowledgments}

The work of A.G.L., J.M.S., M.A.F.S. on the study of deterministic dynamics was supported by the Spanish Ministry of Economy and Competitiveness and
by the Spanish State Research Agency (AEI) and the European Regional
Development Fund (ERDF) under Project No. FIS2016-76883-P. The work of I.B. and L.R. on the study of stochastic dynamics was supported by the Russian Science Foundation (project 16-11-10098).

\nocite{*}

\bibliography{apssamp}

\end{document}